\documentclass[aps,prstab,twocolumn,groupedaddress]{revtex4-1}

\usepackage{graphicx}
\usepackage{subfigure}
\usepackage{comment}
\usepackage{verbatim}
\usepackage{url}

\usepackage{natbib}
\begin{document}


\title{Multi-objective Optimizations of a Normal Conducting RF Gun Based Ultra Fast Electron Diffraction Beamline}


\author{Colwyn Gulliford}\email{cg248@cornell.edu}
\author{Adam Bartnik}
\author{Ivan Bazarov}
\affiliation{CLASSE, Cornell University, 161 Synchrotron Drive
Ithaca, NY 14853-8001}
\author{Jared Maxson}
\affiliation{Department of Physics and Astronomy, University California Los Angeles, Los Angeles, California, 90095, USA}


\date{\today}

\begin{abstract}
We present the results of multi-objective genetic algorithm optimizations of a potential single shot ultra fast electron diffraction beamline utilizing a 100 MV/m  1.6 cell normal conducting rf (NCRF) gun, as well as a 9 cell $2\pi/3$ bunching cavity placed between two solenoids.   Optimizations of the transverse projected emittance as a function of bunch charge are presented and discussed in terms of the scaling laws derived in the charge saturation limit.  Additionally, optimization of the transverse coherence length as a function of final rms bunch length at the sample location have been performed for a charge of $10^6$ electrons.  Analysis of the solutions is discussed, as are the effects of disorder induced heating.  In particular, for a charge of $10^6$ electrons and final beam size of $\sigma_x\geq25$ $\mu$m, we found a relative coherence length of $L_{c,x}/\sigma_x\approx0.07$, $0.1$, and $0.2$ nm/$\mu$m for a final bunch length of $\sigma_t\approx 5$, 30, and 100 fs, respectively.  These results demonstrate the viability of using genetic algorithms in the design and operation of ultrafast electron diffraction beamlines.

\end{abstract}

\pacs{PACS numbers?}


\maketitle

\section{Introduction}
  
The desire for single-shot ultrafast electron diffraction (UED) beamlines ($\sigma_t \lesssim$ 30 fs, $q\sim 10^6$ electrons) capable of imaging molecular and atomic motion continues to push the development of both photocathode and cold atom electron sources \cite{ref:uedUT1,ref:ued:dcbun2solconcept,ref:uedUT2,ref:ued:dcbun2sol2,ref:ued:dcbun2sol,ref:uedUT3,ref:coldatoms1,ref:coldatoms2}.  In the case of photoemission sources, advances in the development of low mean transverse energy (MTE) photocathodes \cite{ref:redmte,ref:coldcathode}, as well as both DC gun and normal conducting rf gun technology \cite{ref:pietro0} now bring the goal of creating single shot electron diffraction beamlines with lengths on the order of meters with in reach.

For such devices, the required charge and beam sizes at the cathode imply transporting a space charged dominated beam.  Building on the successful application of Multi-Objective Genetic Algorithm (MOGA) optimized simulations of space charge dominated beams used in the design and operation of the Cornell photoinjector \cite{ref:lowemitter,ref:lowemitter2,ref:lowemitter3}, as well as the optimization of a cryo-cooled dc gun UED set-up \cite{ref:coldgun}, we apply the same techniques to a 100 MV/m 1.6 cell ncrf gun followed by a 9 cell, $2\pi/3$ buncher cavity.  We use a MTE value of 35 meV, a value considered achievable through the use of multi-alkolide photocathodes operated near threshold \cite{ref:redmte}. 

This work is structured as follows: first, we briefly review the definition of coherence and the expected scaling with critical initial laser and beam parameters.  Next, a detailed description of the beamline set-up, and the parameters for optimization is given.  The results of an initial round of optimizations of the emittance vs. bunch charge, as well as detailed optimizations of the coherence length vs. final bunch length for a charge of $10^6$ electrons follow.  From the optimal fronts, an example simulation for $\sigma_t\approx 5$ fs is presented. 

Throughout this work all fields and particle distributions are assumed symmetric about the beam line ($z$) axis.  For a beam passing through a waist the coherence length expression reduces to \cite{ref:ued:dcbun2solconcept,ref:coldatoms1}
\begin{eqnarray} 
\left.\frac{L_{c,x}}{\lambdabar_e}\right|_{\mathrm{waist}} = \frac{\sigma_x}{\epsilon_{n,x}}.
\label{eqn:lcxemit}
\end{eqnarray} 
where $\lambdabar_e\equiv\hbar/m_ec =3.862...\times10^{-4}$ nm is the \emph{reduced} Compton wavelength of the electron.   For a bunch charge $q$, we write the minimum allowed initial laser spot size in the charge saturation limit for the two limiting cases of a short and long initial laser pulse, as defined by the aspect ratio of $A = \sigma_{x,i} / \Delta z \approx \sigma_{x,i} /  \frac{eE_0}{m_ec^2}(c\sigma_{t,i})^2$. In this expression $E_0$ is the accelerating field at the cathode and $\sigma_{t,i}$ is the laser pulse length \cite{ref:maxbb,ref:pietro2}. The minimum laser spot size is then given by:
\begin{eqnarray}
\sigma_{x,i}\propto
\left\{ \begin{array}{c}
 (q/E_0)^{1/2}, \hspace{0.1cm} A \gg 1 \hspace{0.1cm}(``\mathrm{pancake}")\\
(q/\sigma_{t,i})^{2/3}E_0^{-1}, \hspace{0.1cm} A\leq 1 \hspace{0.1cm} (\mathrm{``cigar"})
\end{array} \right.
\label{eqn:stdxscale}
\end{eqnarray} 
Plugging this into the expression for the cathode emittance $\epsilon_{n,x,i}=\sigma_{x,i}\sqrt{MTE/mc^2}$, where $MTE$ is the mean transverse energy of the photoemitted eletrons, and combing that with Eq.~(\ref{eqn:lcxemit}) yields
\begin{eqnarray}
\frac{L_{c,x}}{\lambdabar_e} \propto  f_{\epsilon}\sigma_x \sqrt{\frac{m_ec^2}{MTE}}.
\left\{ \begin{array}{c}
 (E_0/q)^{1/2}, \hspace{0.1cm} A \gg 1 \\
E_0(\sigma_{t,i}/q)^{2/3}, \hspace{0.1cm} A\lesssim 1 
\end{array}. \right.
\label{eqn:lcxscale}
\end{eqnarray}
Here the factor $f_{\epsilon} = \epsilon_{n,x,i}/\epsilon_{n,x,f}$ is included because the above expression assumes the emittance at the cathode can be recovered at the sample.  For beams with a large degree of emittance preservation, $f_{\epsilon}\approx 1$, and the above expression gives the correct scaling \cite{ref:maxbb,ref:pietro2}.
 
As previously mentioned, the NCRF gun based UED setup consists of a 1.6 cell 2.856 GHz NCRF gun followed by a buncher cavity placed between two solenoid magnets \cite{ref:pietro1,ref:pietro2,ref:rfsimpap}.  For all simluations, the maximum allowable peak cathode field is set to 100 MV/m and the cathode MTE is fixed to 35 meV for all simulations.  
For the buncher model, we used the dimensions of the first cell in the SLAC linac, specified in Table 6-6 in \cite{ref:slaccav}, repeated a total of 9 times, to make a 9-cell $2 \pi/3$ traveling wave buncher cavity. The fieldmaps were generated in Poisson Superfish.  To model solenoids similar to those in \cite{ref:pietro1,ref:pietro2} we use an analytic form for the on-axis solenoid field from a sheet of current with radius $R$ and length $L$,
\begin{eqnarray}
B_z(z) = B_0\left( \frac{\Delta z_+}{\sqrt{\Delta z_+^2 + R^2}} -  \frac{\Delta z_-}{\sqrt{\Delta z_-^2 + R^2}}\right), 
\label{eqn:solfield}
\end{eqnarray}
where $\Delta z_{\pm} =  z \pm L/2$, to the solenoid field maps, and created a custom GPT element featuring the analytic result of the off-axis expansion of Eq.~(\ref{eqn:solfield}) to third order  in the radial offset $r$.  Note that given the small beam sizes along each set-up ($\sigma_x \lesssim 1$ mm) determined by the optimizer, the first-order expansion of the solenoid fields accurately describes the beam dynamics.  
 \begin{figure}[ht!]
    \begin{center}
        \subfigure[\hspace{0.2cm}Example of the on-axis accelerating and solenoid field profiles.]{%
           \label{fig:cgunlayout}
           \includegraphics[width=0.45\textwidth]{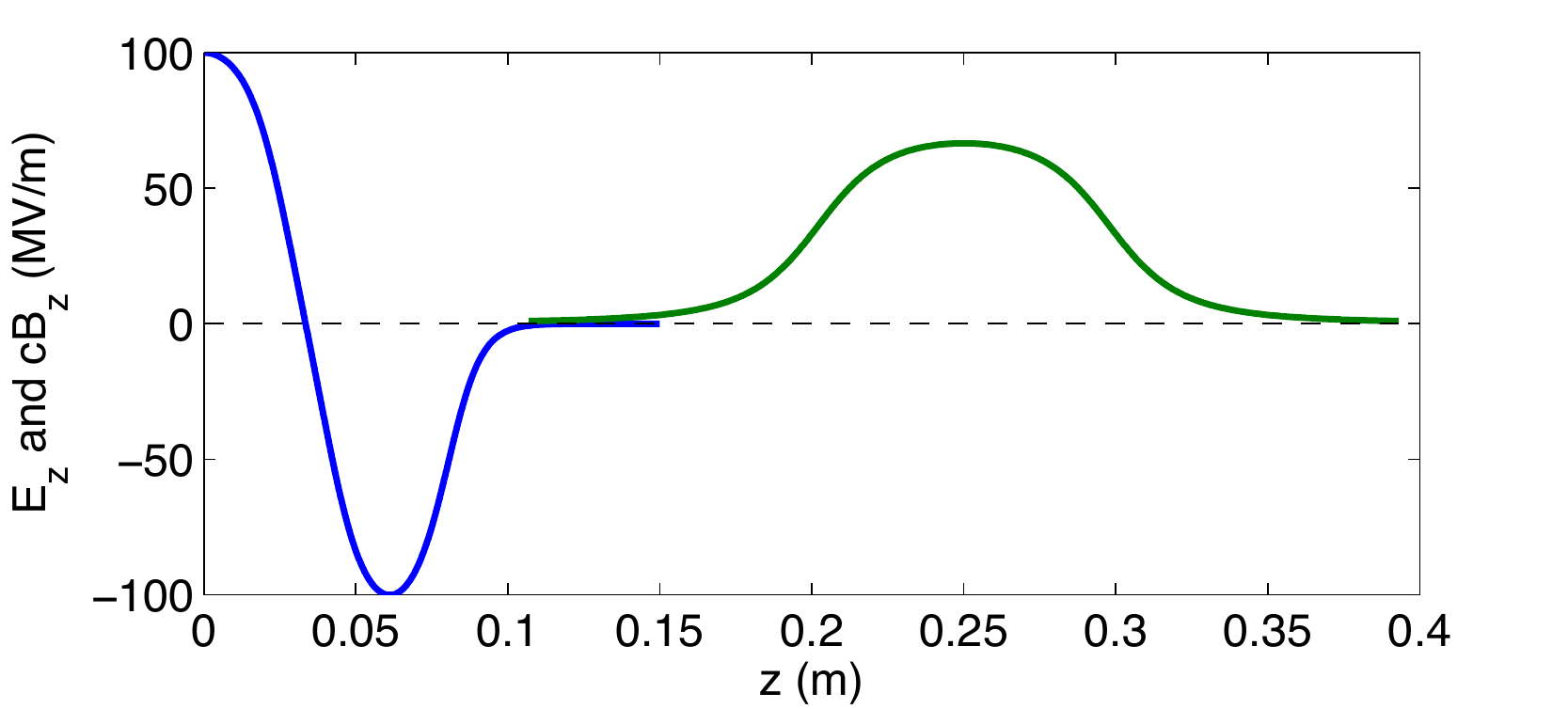}
        }\\ 
        \subfigure[\hspace{0.2cm}The 9 cell buncher cavity design and fields for an input peak power of 20 MW.]{%
           \label{fig:ncrflayout}
           \includegraphics[width=0.45\textwidth]{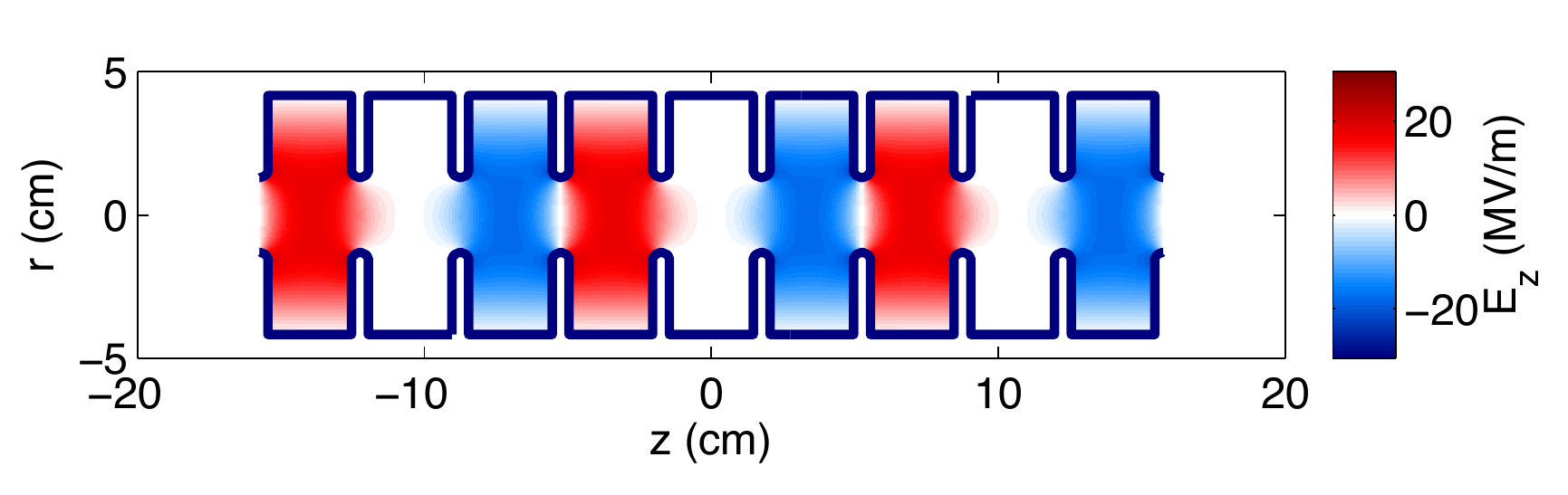}
        }\\ 
    \end{center}
    \caption{%
        \label{fig:3Dslice}
    Field description for the 1.6 cell gun and solenoid (a), as well as the buncher (b).}%
\end{figure}
Additionally, the use of such small MTE and laser spot size values requires estimating the effect of disorder induced heating (DIH) near the cathode.  This issue is discussed later in the results section.

\section{Results}

In order to produce the best coherence length performance from the UED setup, multi-objective genetic optimizations were performed using General Particle Tracer and the same optimization software used previously in \cite{ref:lowemitter,ref:lowemitter2,ref:lowemitter3,ref:coldgun}.  For these simulations, the optimizer varied the laser rms sizes, beamline element parameters and positions.  Additionally, the optimizer was allowed to arbitrarily shape both the transverse and longitudinal laser distributions, based on the same method described in \cite{ref:dcrfcomp}.  Table-\ref{tab:beamline_params} displays the beamline parameters varied.
\begin{table}[htb]
\caption{Beamline Simulation Parameters}\label{tab:beamline_params}
\begin{ruledtabular}
\begin{tabular}{ cc | cc }
Parameter & & Range  \\ 
\hline
Laser Size $\sigma_{t,i}$ &&  [0, 20] ps \\
Laser Size $\sigma_{x,i}$ && [0, 1] mm  \\
Cathode MTE && 35 meV  \\
Peak Gun Field && 100 MV/m \\
Gun Phase && [0, 360] deg \\
Solenoid Radius/Length && 2.53/9.54 cm\\
Solenoid Peak Field(s) && [0, 1.0] T\\
Solenoid 1 Position && [0.25,  5.25]  m\\
Peak Buncher Power && [0, 25] MW \\%
Buncher Phase && [0, 360] deg \\
Buncher Position && [0.4, 10.4] m \\
Solenoid 2 Position && [0.55, 15.55]  m\\
Sample Position && [0.55 20.55] m
\end{tabular}
\end{ruledtabular}
\end{table}

\subsubsection{Optimal Emittance}

\begin{figure}[h!]
 \centering
\includegraphics[width=0.45\textwidth]{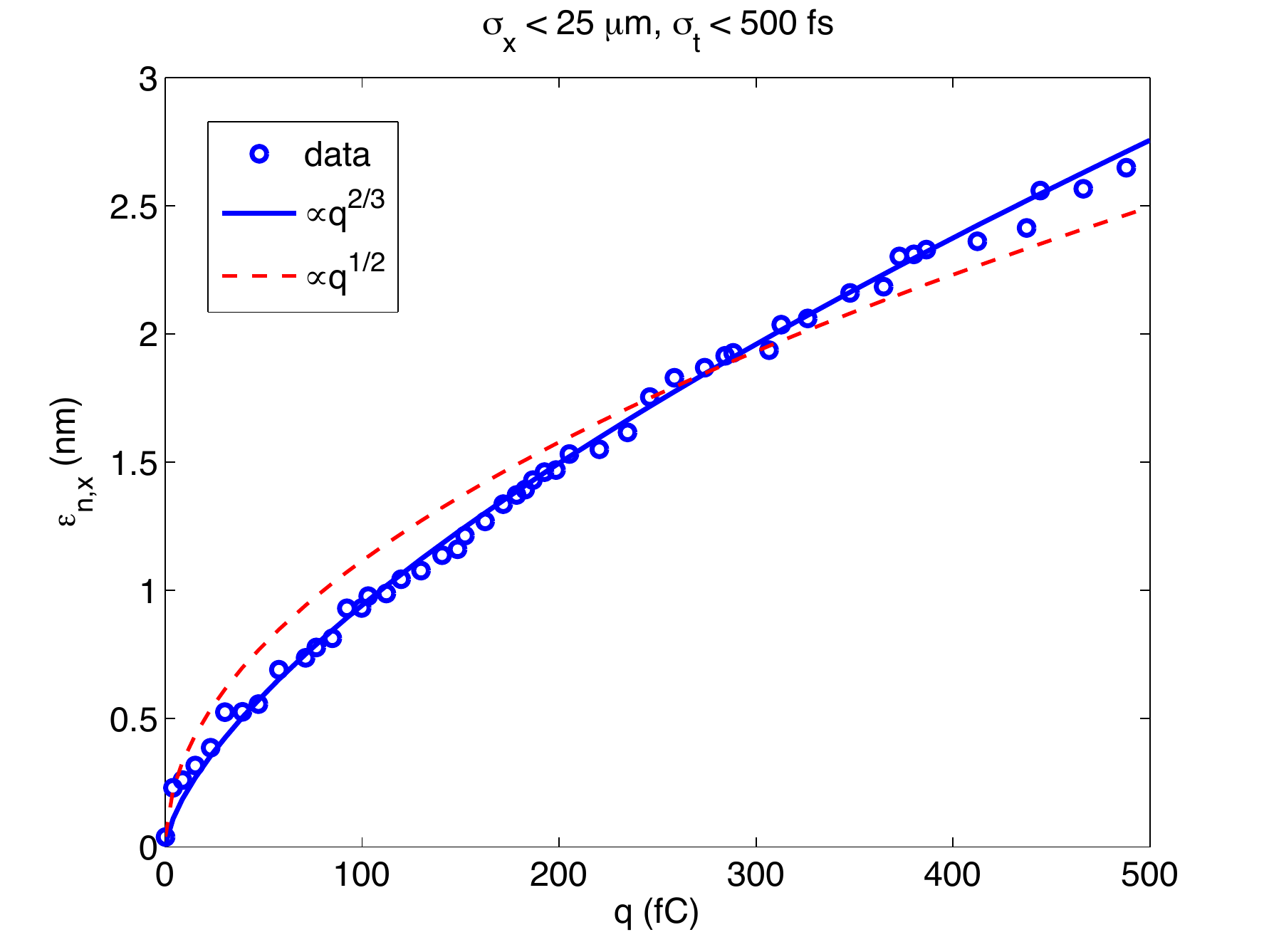}
 \caption{Optimal emittance as a function of bunch charge at the sample for final beam size of $\sigma_x\leq25$, $\sigma\leq500$ fs.}
\label{fig:enx_vs_q}
\end{figure}

Given a final spot size $\sigma_x$ Eqn.~(\ref{eqn:lcxemit}) implies the fundamental limit to the coherence is the emittance at the sample.   As previously stated, the emittance preservation factor $f_{\epsilon}$ in Eqn.~(\ref{eqn:lcxscale}) determines the degree to which the scaling laws in Eqn.~(\ref{eqn:lcxscale}) hold true, and may depend strongly on both the initial and final beam sizes.  To determine the effects of constraining the final required rms sizes, we perform an initial round of optimizations for a ``long" final beam, $\sigma_x\leq25$ $\mu$m and $\sigma_t\leq 500$ fs.  In these optimizations, we require that no particles are lost in beam transport.  Fig.~\ref{fig:enx_vs_q} shows the emittance performance.  
The figure also shows the results of fitting the emittance data with both scaling laws found in Eqn.~(\ref{eqn:stdxscale}), and suggests that the emittance at the sample does scale as $q^{2/3}$.  Noting that the optimizer chose an initial laser pulse length of roughly 2.4 ps for the charges shown, and computing the initial beam aspect ratio for all points on the front leads to an aspect ratio range of roughly 0.006 to 0.1, verifying the fitted scaling law.
 
\begin{figure}[h!]
 \centering
\includegraphics[width=0.45\textwidth]{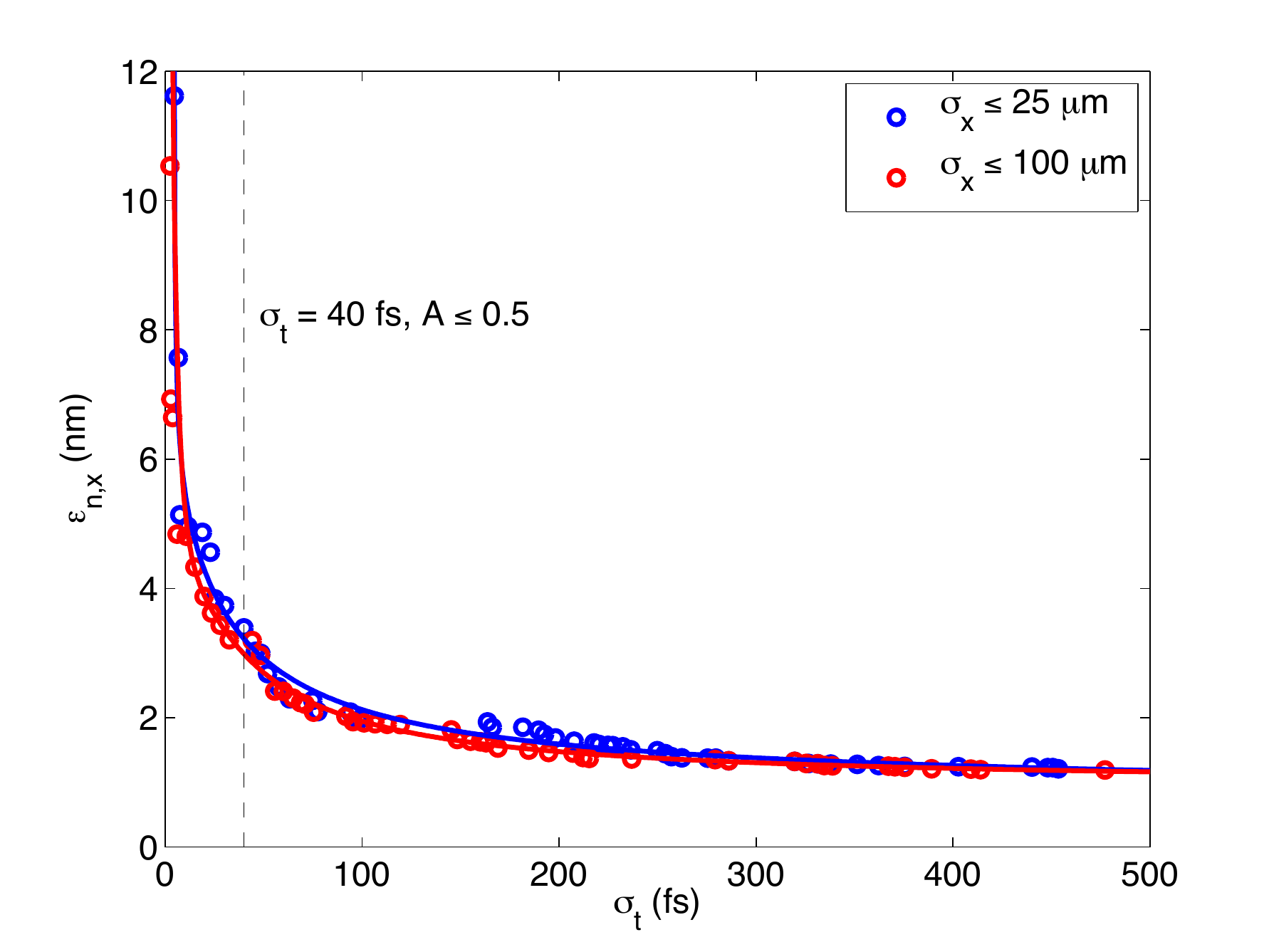}
 \caption{The optimal emittances for a final beam size of $\sigma_x\leq25$ $\mu$m as a function of bunch length.}
\label{fig:EvsT}
\end{figure}

Taking these results, we fixed the charge to $10^6$ electrons and re-optimized, this time minimizing the emittance as a function of bunch length.  Fig.~\ref{fig:EvsT} shows the results for two final beam spot sizes at the sample: $\sigma_x\leq25$ and 100 $\mu$m and demonstrates that the emittance scales weakly with spot size for final beam sizes of $\sigma_x\geq25$ $\mu$m.  Additionally, the data shows the emittance scales weakly with the final bunch length down to roughly $\sigma_t\sim100$ fs.  For shorter bunch lengths the emittance suffers, though we note the existence of feasible solutions with small emittances $\epsilon_{n,x}\lesssim5$ nm for bunch lengths as short as $\sigma_t\sim5$ fs.  Note, for this plot and all similar ones, we fit a rational polynomial to the Pareto front in order to better guide the eye and to aid estimating and interpolating between points on the front.  
The dashed black line in this plot indicates where the optimizer selected initial rms laser sizes corresponding to an initial beam aspect ratio of $A\approx0.5$.  Beams to the right of this line are considered to be created in the long pulse regime.  It is instructive to also estimate the disordered induced heating of the beam for this data.  To do so we assume a uniform beam with equivalent rms sizes.  From this, the volume of the beam after being emitted from the cathode follows:
\begin{eqnarray}
V &=& \pi R^2 L\approx \pi (2\sigma_x)^2 \cdot \frac{1}{2}\frac{eE_0}{m} (\sqrt{12}\sigma_t)^2 
\nonumber\\
&\approx& \frac{24\pi E_0}{mc^2 [\mathrm{eV}]}\sigma_x^2(c\sigma_t)^2
\end{eqnarray}
From this we estimate the effect of disordered induced heating using the formula given by Maxson: $\Delta kT \hspace{0.1cm}[\mathrm{eV}]  = 1.04\times10^{-9}(n_0 \hspace{0.1cm}[m^{-3}])^{1/3}$ \cite{ref:DIH}, where $n_0$ is the electron number density.  For the shortest final bunch lengths, the DIH estimate is roughly 150 meV.  It is important to note that this estimate may indicate that DIH will be important, however, the true estimate of the effect depends on the detailed dynamics of the initial expansion of the beam, and lies beyond the scope of this work.


\subsubsection{Optimal Coherence Length}
Using the emittance vs. bunch length solutions for a final spot size of 25 $\mu$m in Fig.~\ref{fig:EvsT} as a seed, optimizations of the transverse coherence length were performed.
 \begin{figure}[htb!]
\centering
\includegraphics[width=0.42\textwidth]{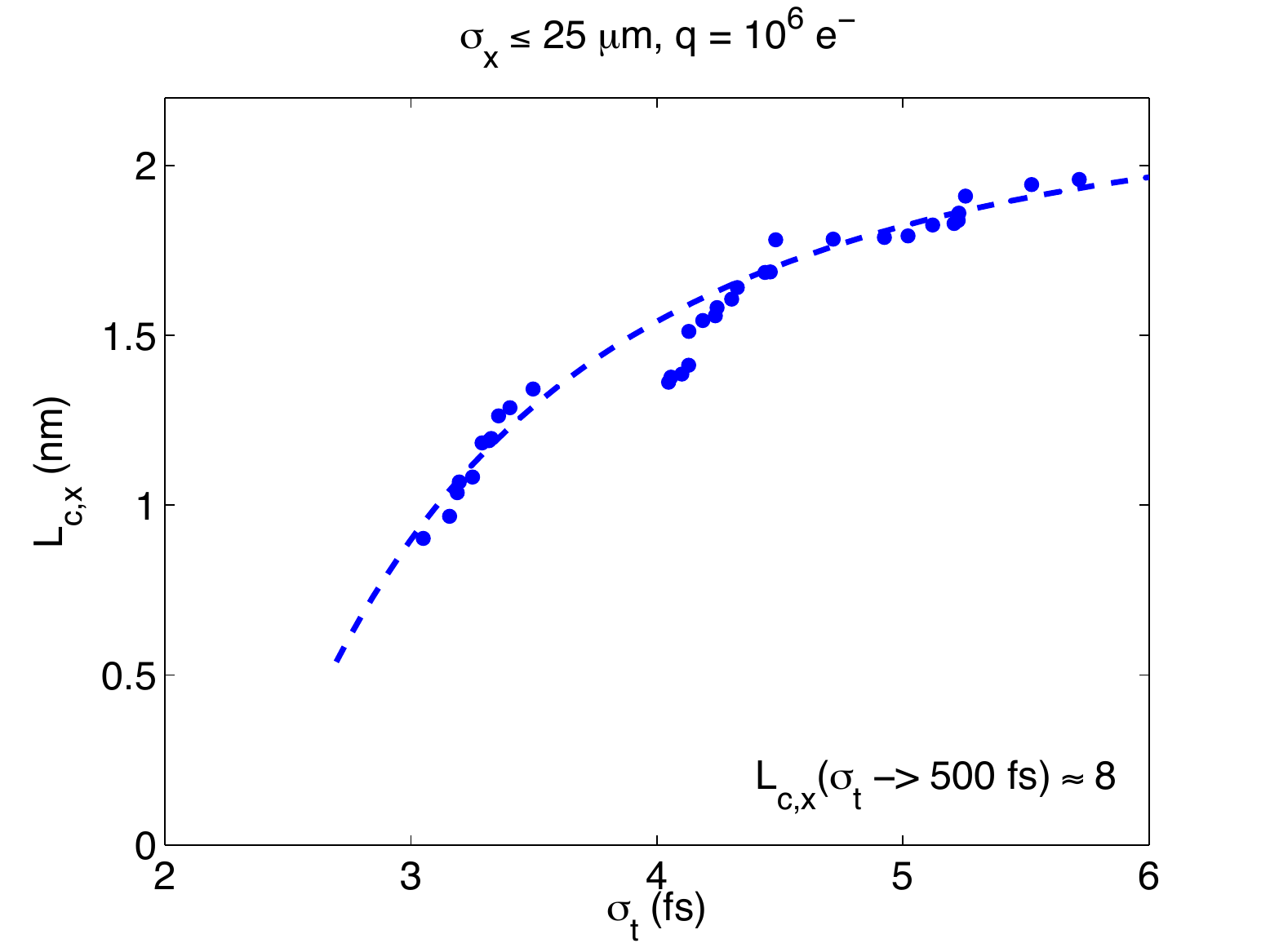}
\caption{Optimal coherence length as a function of bunch length length at the sample for a final beam spot size of $\sigma_z\leq25$ $\mu$m.}
\label{fig:eLvsT}
\end{figure}
Fig.~\ref{fig:eLvsT} shows the optimal coherence length as a function of final bunch length $\sigma_t$, constrained so that $\sigma_x\leq25$ $\mu$m.  
\begin{figure}[h]
\centering
\includegraphics[width=0.42\textwidth]{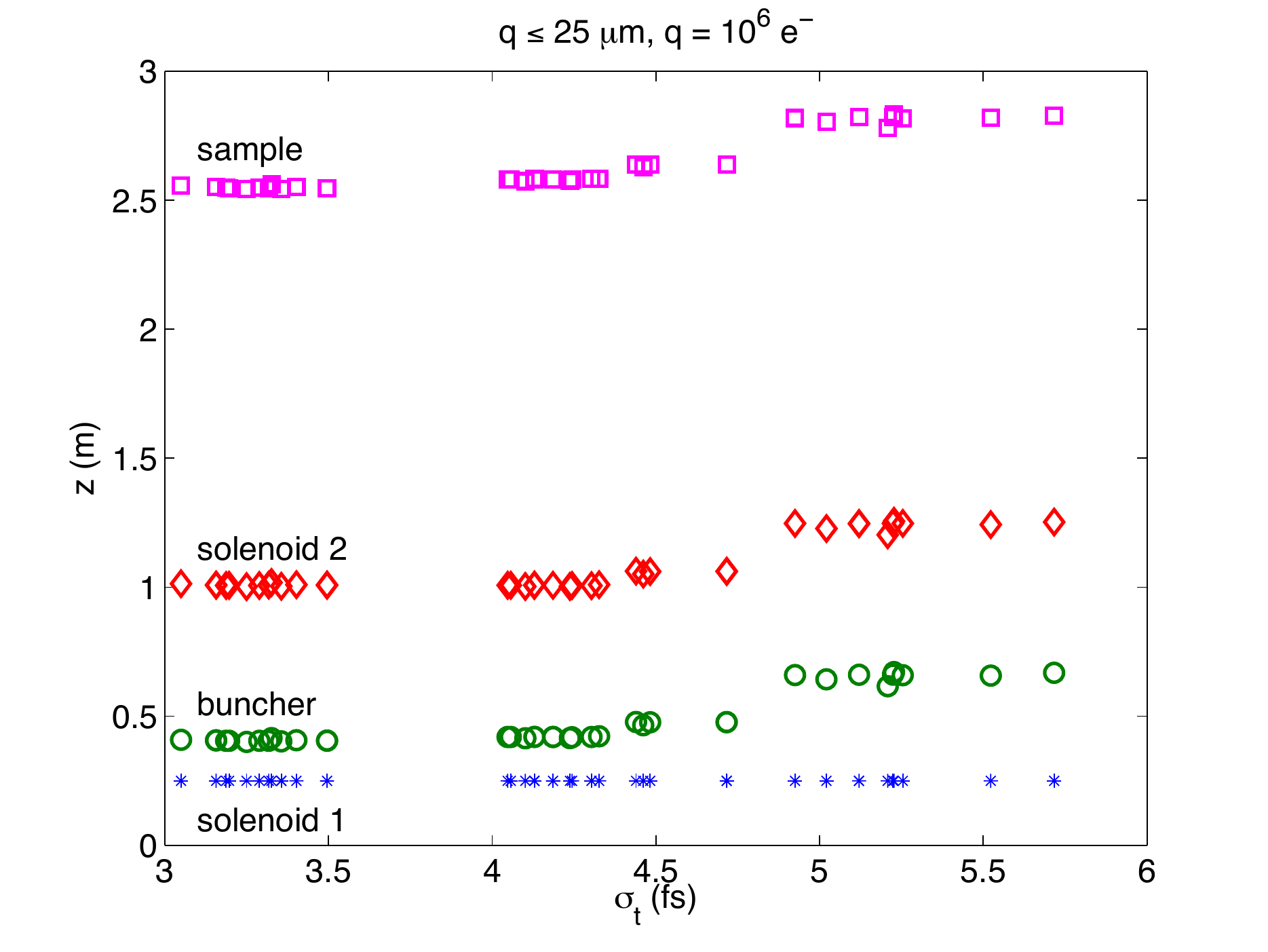}
\caption{Optimal beamline element positions for a final beam spot size of $\sigma_z\leq25$ $\mu$m.}
\label{fig:eLvsT}
\end{figure}
In particular, the data show relative coherence lengths of $L_{c,x}/\sigma_x\approx0.07$, $0.1$, and $0.2$ nm/$\mu$m for a final bunch length of $\sigma_t\approx 5$, 30, and 100 fs, respectively.  We estimate the limiting value for the coherence length for the case of a long beam (500 fs) at the sample to be roughly 8 nm, using the emittance data from Fig.~\ref{fig:EvsT} and Eqn.~(\ref{eqn:lcxemit}).  Barring any effect from DIH at the shortest bunch lengths, these results demonstrate the viability of using genetic algorithms in the design and operation of ultrafast electron diffraction beamlines.  We compare these results to those found in \cite{ref:rfsimpap}.  At roughly the same bunch length (5 fs), the relative coherence from Jang is about 0.014 nm/$\mu $m, implying a factor of 5 improvement in our results.  We note, however, that an MTE of 600 meV and a max accelerating field of 80 MV/m were used in \cite{ref:rfsimpap}.  Accounting for these differences would bring the relative coherence length in \cite{ref:rfsimpap} up to 0.07, in agreement with this work.  Thus, we point out the majority of the improvement comes from using better cathode. 

In addition to determining the the optimal coherence length, the optimizations producing the data in Fig.~\ref{fig:eLvsT} also provides information about the optimal positioning of the beamline elements in each set-up.
Table-\ref{tab:beamline_params} displays the element positions averaged over the optimization data shown in Fig.~\ref{fig:eLvsT}.
\begin{table}[htb]
\caption{Average Optimized Beamline Element Positions}\label{tab:beamline_params}
\begin{ruledtabular}
\begin{tabular}{ cc | cc  }
Element &&  Position  \\ 
\hline
Solenoid 1  && 0.25 m   \\
Buncher Cavity && 0.48 m \\
Solenoid 2 &&  1.07 m \\
Sample Pinhole  && 2.63 m   
\end{tabular}
\end{ruledtabular}
\end{table}

\subsubsection{Example Solution}
In order to get a better feel of the beam dynamics determined by the coherence length optimizations, we ran a single example solution from the coherence vs. final bunch length fronts shown in Fig.~\ref{fig:eLvsT}.  This solution has a kinetic energy of 4.5 MeV, typical of all of the simulations presented in this work.
Table-\ref{tab:resparams} displays the resulting relevant beam parameters.  Of note are the initial laser sizes of $\sigma_{x,i}\approx 5$ $\mu$m and $\sigma_{t,i}\approx32$ fs.  These correspond to an initial beam aspect ratio of roughly $4\times10^4$ and an estimated DIH effect of 176 meV, a value which may indicate that the space charge model is not sufficient to describe the dynamics near the cathode \cite{ref:DIH}.  We note that though the optimizer did not generate solutions here at longer bunch lengths, the optimal emittance results suggest viable solutions exist a longer bunch lengths $\sigma_t \approx 40$ fs.
\begin{table}[h!]
\caption{Example parameters and results.}\label{tab:resparams}
\begin{ruledtabular}
\begin{tabular}{ c  c}
Parameter & Simulated Value\\ 
\hline
q & $10^6$ electrons \\
Gun Phase (from peak field) & -47 deg \\
Buncher Phase (from on-crest) & -90 \\
Buncher Peak Power & 22.5 MW\\
Laser $\sigma_{x,i}$ & 4.68 $\mu$m  \\
Laser $\sigma_{t,i}$ & 32 fs \\
Aspect Ratio A & $4\times10^4$ \\
KE & 4.5 MeV\\
$\sigma_x$ & $\leq$ 25 $\mu$m\\
$\sigma_t$ & 5 fs \\
$\epsilon_{n,x}$ & 5.2 nm\\
$L_{c,x}$ & 1.85 nm\\
\end{tabular}
\end{ruledtabular}
\end{table}
Fig.\ref{fig:stdx} shows the transverse rms beamsize along the beamline, as well as the initial transverse laser profile and the final electron transverse distribution at the sample.  The optimizer chose a roughly flattop transverse laser profile with $\sigma_x\approx 5$ $\mu$m.  
\begin{figure*}[ht]
    \begin{center}
        \subfigure[\hspace{0.2cm}]{%
           \label{fig:stdx}
           \includegraphics[width=0.95\textwidth]{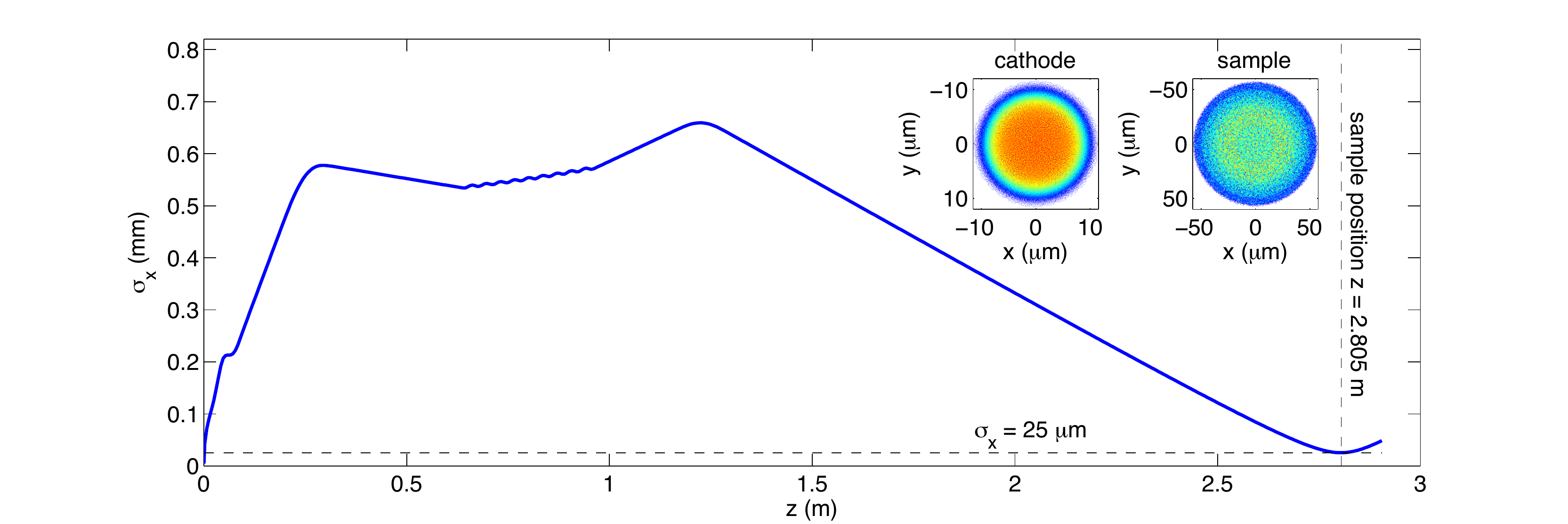}
        }\\ 
        \subfigure[\hspace{0.2cm}]{%
           \label{fig:stdt}
           \includegraphics[width=0.95\textwidth]{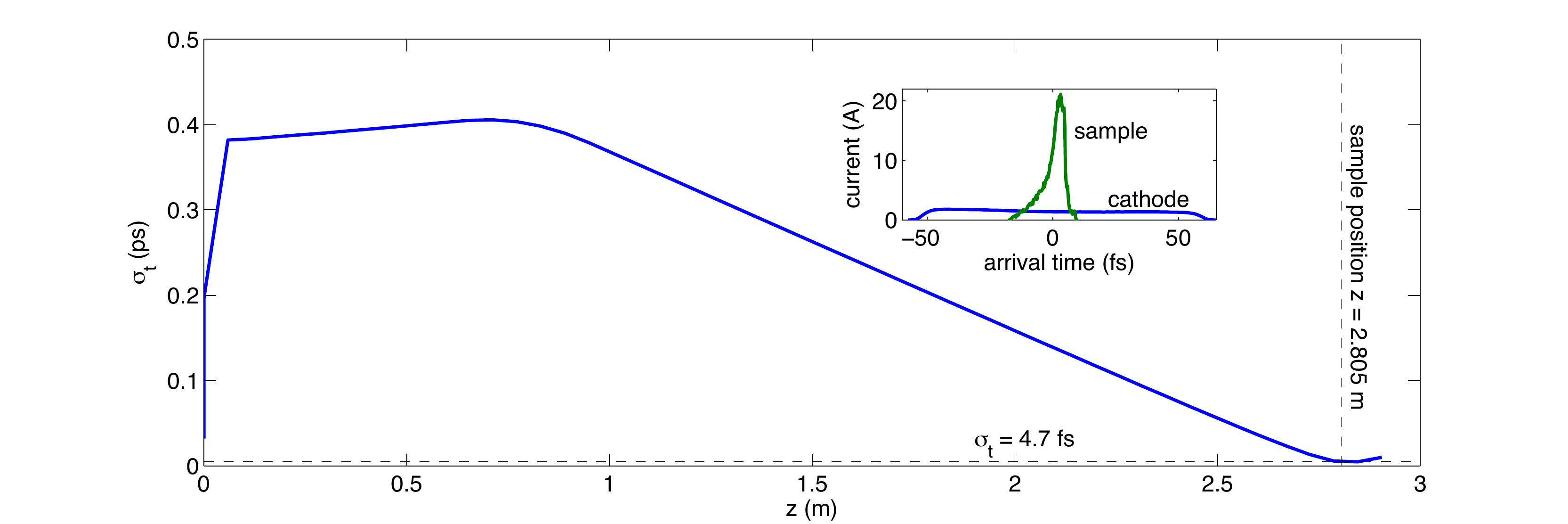}
        }\\ 
        \subfigure[\hspace{0.2cm}]{%
           \label{fig:enx}
           \includegraphics[width=0.95\textwidth]{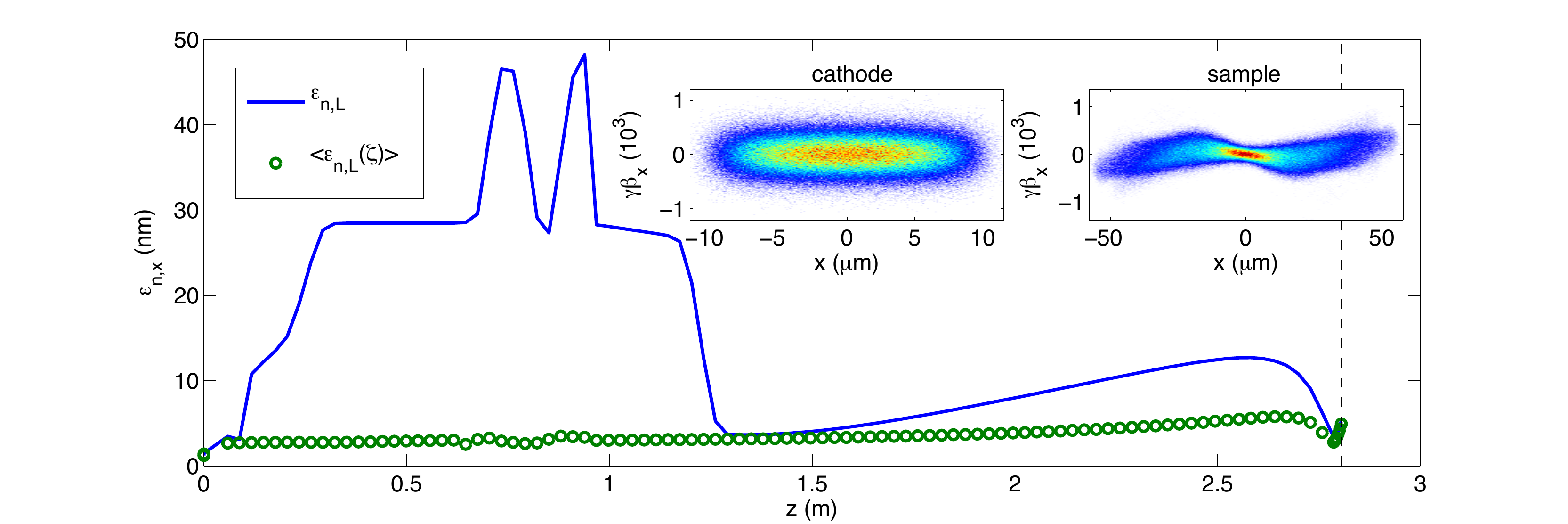}
        }\\ 
    \end{center}
    \caption{%
    \label{fig:stdx}
    Transverse (a) and longitudinal (b) rms beam size along the beamline, as well as the transverse emittance (c).  Insets show the transverse (a) and longitudinal (b) beam distributions and (c) transverse phase spaces at the cathode and sample locations for both final charges.}%
\end{figure*}
Fig.~\ref{fig:stdt} shows the rms bunch length, and the initial temporal current profile produced by the laser, and the electron beam current profile at the sample.  The use of the buncher cavity allows for a fairly constant bunch length along the beamline up to the cavity, where the buncher applies an energy chirp which results in the bunch being compressed by the time it reaches the screen.  The relevant emittance data along the beamline is shown in Fig.~\ref{fig:enx}.  Shown in solid blue is the transverse emittance with the angular velocity due to the solenoid removed.  Shown in green is the slice emittance, computed by averaging over the (30) individual slices.  As anticipated, the individual slice phase spaces are nearly aligned at the sample.

In this work, we have presented a multi-MeV NCRF gun based single shot UED layout determined by MOGA optimization of space charge simulations.  For a long final bunch of 500 fs or less, the emittance for this set-up scales as $q^{2/3}$, in agreement with the predictions for a long initial beam aspect ratio \cite{ref:pietro1,ref:pietro2}.  Emittances as low as 2-5 nm were found for final bunch lengths ranging from roughly 500 down to 5 fs for a bunch charge of $10^6$ electrons.  Estimates of the DIH and initial beam aspect ratio show no issues for bunch lengths as small as 40 fs.  In addition to computing the optimal emittances as a function of bunch length, optimizations of the coherence length as a function final bunch length produced coherence lengths suitable for single-shot UED experiments with a final electron beam spot size of $\sigma_x\geq25$ and bunch charge of $q\leq10^6$ electrons.  In particular, optimal relative coherence lengths of $L_{c,x}/\sigma_x\approx0.07$, $0.1$, and $0.2$ nm/$\mu$m for a final bunch length of $\sigma_t\approx 5$, 30, and 100 fs, respectively, were produced.  


\begin{acknowledgments}
This grant was supported by the NSF, Award PHY 1416318.   
\end{acknowledgments}


\begin{thebibliography}{22}%
\makeatletter
\providecommand \@ifxundefined [1]{%
 \@ifx{#1\undefined}
}%
\providecommand \@ifnum [1]{%
 \ifnum #1\expandafter \@firstoftwo
 \else \expandafter \@secondoftwo
 \fi
}%
\providecommand \@ifx [1]{%
 \ifx #1\expandafter \@firstoftwo
 \else \expandafter \@secondoftwo
 \fi
}%
\providecommand \natexlab [1]{#1}%
\providecommand \enquote  [1]{``#1''}%
\providecommand \bibnamefont  [1]{#1}%
\providecommand \bibfnamefont [1]{#1}%
\providecommand \citenamefont [1]{#1}%
\providecommand \href@noop [0]{\@secondoftwo}%
\providecommand \href [0]{\begingroup \@sanitize@url \@href}%
\providecommand \@href[1]{\@@startlink{#1}\@@href}%
\providecommand \@@href[1]{\endgroup#1\@@endlink}%
\providecommand \@sanitize@url [0]{\catcode `\\12\catcode `\$12\catcode
  `\&12\catcode `\#12\catcode `\^12\catcode `\_12\catcode `\%12\relax}%
\providecommand \@@startlink[1]{}%
\providecommand \@@endlink[0]{}%
\providecommand \url  [0]{\begingroup\@sanitize@url \@url }%
\providecommand \@url [1]{\endgroup\@href {#1}{\urlprefix }}%
\providecommand \urlprefix  [0]{URL }%
\providecommand \Eprint [0]{\href }%
\providecommand \doibase [0]{http://dx.doi.org/}%
\providecommand \selectlanguage [0]{\@gobble}%
\providecommand \bibinfo  [0]{\@secondoftwo}%
\providecommand \bibfield  [0]{\@secondoftwo}%
\providecommand \translation [1]{[#1]}%
\providecommand \BibitemOpen [0]{}%
\providecommand \bibitemStop [0]{}%
\providecommand \bibitemNoStop [0]{.\EOS\space}%
\providecommand \EOS [0]{\spacefactor3000\relax}%
\providecommand \BibitemShut  [1]{\csname bibitem#1\endcsname}%
\let\auto@bib@innerbib\@empty
\bibitem [{\citenamefont {Siwick}\ \emph {et~al.}(2004)\citenamefont {Siwick},
  \citenamefont {Dwyer}, \citenamefont {Jordan},\ and\ \citenamefont
  {Miller}}]{ref:uedUT1}%
  \BibitemOpen
  \bibfield  {author} {\bibinfo {author} {\bibfnamefont {B.}~\bibnamefont
  {Siwick}}, \bibinfo {author} {\bibfnamefont {J.}~\bibnamefont {Dwyer}},
  \bibinfo {author} {\bibfnamefont {R.}~\bibnamefont {Jordan}}, \ and\ \bibinfo
  {author} {\bibfnamefont {R.}~\bibnamefont {Miller}},\ }\href {\doibase
  http://dx.doi.org/10.1016/j.chemphys.2003.11.040} {\bibfield  {journal}
  {\bibinfo  {journal} {Chemical Physics}\ }\textbf {\bibinfo {volume} {299}},\
  \bibinfo {pages} {285 } (\bibinfo {year} {2004})},\ \bibinfo {note}
  {ultrafast Science with X-rays and Electrons}\BibitemShut {NoStop}%
\bibitem [{\citenamefont {van Oudheusden}\ \emph {et~al.}(2007)\citenamefont
  {van Oudheusden}, \citenamefont {de~Jong}, \citenamefont {van~der Geer},
  \citenamefont {'t~Root}, \citenamefont {Luiten},\ and\ \citenamefont
  {Siwick}}]{ref:ued:dcbun2solconcept}%
  \BibitemOpen
  \bibfield  {author} {\bibinfo {author} {\bibfnamefont {T.}~\bibnamefont {van
  Oudheusden}}, \bibinfo {author} {\bibfnamefont {E.~F.}\ \bibnamefont
  {de~Jong}}, \bibinfo {author} {\bibfnamefont {S.~B.}\ \bibnamefont {van~der
  Geer}}, \bibinfo {author} {\bibfnamefont {W.~P. E. M.~O.}\ \bibnamefont
  {'t~Root}}, \bibinfo {author} {\bibfnamefont {O.~J.}\ \bibnamefont {Luiten}},
  \ and\ \bibinfo {author} {\bibfnamefont {B.~J.}\ \bibnamefont {Siwick}},\
  }\href {\doibase http://dx.doi.org/10.1063/1.2801027} {\bibfield  {journal}
  {\bibinfo  {journal} {Journal of Applied Physics}\ }\textbf {\bibinfo
  {volume} {102}},\ \bibinfo {eid} {093501} (\bibinfo {year}
  {2007})}\BibitemShut {NoStop}%
\bibitem [{\citenamefont {Harb}(2009)}]{ref:uedUT2}%
  \BibitemOpen
  \bibfield  {author} {\bibinfo {author} {\bibfnamefont {M.}~\bibnamefont
  {Harb}},\ }\emph {\bibinfo {title} {Investigating Photoinduced Structural
  Changes in Si using Femtosecond Electron Diffraction}},\ \href@noop {} {Ph.D.
  thesis},\ \bibinfo  {school} {University of Toronto} (\bibinfo {year}
  {2009})\BibitemShut {NoStop}%
\bibitem [{\citenamefont {van Oudheusden}\ \emph {et~al.}(2010)\citenamefont
  {van Oudheusden}, \citenamefont {Pasmans}, \citenamefont {van~der Geer},
  \citenamefont {de~Loos}, \citenamefont {van~der Wiel},\ and\ \citenamefont
  {Luiten}}]{ref:ued:dcbun2sol2}%
  \BibitemOpen
  \bibfield  {author} {\bibinfo {author} {\bibfnamefont {T.}~\bibnamefont {van
  Oudheusden}}, \bibinfo {author} {\bibfnamefont {P.~L. E.~M.}\ \bibnamefont
  {Pasmans}}, \bibinfo {author} {\bibfnamefont {S.~B.}\ \bibnamefont {van~der
  Geer}}, \bibinfo {author} {\bibfnamefont {M.~J.}\ \bibnamefont {de~Loos}},
  \bibinfo {author} {\bibfnamefont {M.~J.}\ \bibnamefont {van~der Wiel}}, \
  and\ \bibinfo {author} {\bibfnamefont {O.~J.}\ \bibnamefont {Luiten}},\
  }\href {\doibase 10.1103/PhysRevLett.105.264801} {\bibfield  {journal}
  {\bibinfo  {journal} {Phys. Rev. Lett.}\ }\textbf {\bibinfo {volume} {105}},\
  \bibinfo {pages} {264801} (\bibinfo {year} {2010})}\BibitemShut {NoStop}%
\bibitem [{\citenamefont {Chatelain}\ \emph {et~al.}(2012)\citenamefont
  {Chatelain}, \citenamefont {Morrison}, \citenamefont {Godbout},\ and\
  \citenamefont {Siwick}}]{ref:ued:dcbun2sol}%
  \BibitemOpen
  \bibfield  {author} {\bibinfo {author} {\bibfnamefont {R.~P.}\ \bibnamefont
  {Chatelain}}, \bibinfo {author} {\bibfnamefont {V.~R.}\ \bibnamefont
  {Morrison}}, \bibinfo {author} {\bibfnamefont {C.}~\bibnamefont {Godbout}}, \
  and\ \bibinfo {author} {\bibfnamefont {B.~J.}\ \bibnamefont {Siwick}},\
  }\href {\doibase http://dx.doi.org/10.1063/1.4747155} {\bibfield  {journal}
  {\bibinfo  {journal} {Applied Physics Letters}\ }\textbf {\bibinfo {volume}
  {101}},\ \bibinfo {eid} {081901} (\bibinfo {year} {2012})}\BibitemShut
  {NoStop}%
\bibitem [{\citenamefont {Gao}\ \emph {et~al.}(2013)\citenamefont {Gao},
  \citenamefont {Lu}, \citenamefont {Jean-Ruel}, \citenamefont {Liu},
  \citenamefont {Marx}, \citenamefont {Onda}, \citenamefont {Koshihara},
  \citenamefont {Nakano}, \citenamefont {Shao}, \citenamefont {Hiramatsu},
  \citenamefont {Saito}, \citenamefont {Yamochi}, \citenamefont {Cooney},
  \citenamefont {Moriena}, \citenamefont {Sciaini},\ and\ \citenamefont
  {Miller}}]{ref:uedUT3}%
  \BibitemOpen
  \bibfield  {author} {\bibinfo {author} {\bibfnamefont {M.}~\bibnamefont
  {Gao}}, \bibinfo {author} {\bibfnamefont {C.}~\bibnamefont {Lu}}, \bibinfo
  {author} {\bibfnamefont {H.}~\bibnamefont {Jean-Ruel}}, \bibinfo {author}
  {\bibfnamefont {L.~C.}\ \bibnamefont {Liu}}, \bibinfo {author} {\bibfnamefont
  {A.}~\bibnamefont {Marx}}, \bibinfo {author} {\bibfnamefont {K.}~\bibnamefont
  {Onda}}, \bibinfo {author} {\bibfnamefont {S.-y.}\ \bibnamefont {Koshihara}},
  \bibinfo {author} {\bibfnamefont {Y.}~\bibnamefont {Nakano}}, \bibinfo
  {author} {\bibfnamefont {X.}~\bibnamefont {Shao}}, \bibinfo {author}
  {\bibfnamefont {T.}~\bibnamefont {Hiramatsu}}, \bibinfo {author}
  {\bibfnamefont {G.}~\bibnamefont {Saito}}, \bibinfo {author} {\bibfnamefont
  {H.}~\bibnamefont {Yamochi}}, \bibinfo {author} {\bibfnamefont {R.~R.}\
  \bibnamefont {Cooney}}, \bibinfo {author} {\bibfnamefont {G.}~\bibnamefont
  {Moriena}}, \bibinfo {author} {\bibfnamefont {G.}~\bibnamefont {Sciaini}}, \
  and\ \bibinfo {author} {\bibfnamefont {R.~J.~D.}\ \bibnamefont {Miller}},\
  }\href {http://dx.doi.org/10.1038/nature12044} {\bibfield  {journal}
  {\bibinfo  {journal} {Nature}\ }\textbf {\bibinfo {volume} {496}},\ \bibinfo
  {pages} {343} (\bibinfo {year} {2013})}\BibitemShut {NoStop}%
\bibitem [{\citenamefont {Engelen}\ \emph {et~al.}(2013)\citenamefont
  {Engelen}, \citenamefont {van~der Heijden}, \citenamefont {Bakker},
  \citenamefont {Vredenbregt},\ and\ \citenamefont {Luiten}}]{ref:coldatoms1}%
  \BibitemOpen
  \bibfield  {author} {\bibinfo {author} {\bibfnamefont {W.~J.}\ \bibnamefont
  {Engelen}}, \bibinfo {author} {\bibfnamefont {M.~A.}\ \bibnamefont {van~der
  Heijden}}, \bibinfo {author} {\bibfnamefont {D.~J.}\ \bibnamefont {Bakker}},
  \bibinfo {author} {\bibfnamefont {E.~J.~D.}\ \bibnamefont {Vredenbregt}}, \
  and\ \bibinfo {author} {\bibfnamefont {O.~J.}\ \bibnamefont {Luiten}},\
  }\href {http://dx.doi.org/10.1038/ncomms2700} {\bibfield  {journal} {\bibinfo
   {journal} {Nat Commun}\ }\textbf {\bibinfo {volume} {4}},\ \bibinfo {pages}
  {1693} (\bibinfo {year} {2013})}\BibitemShut {NoStop}%
\bibitem [{\citenamefont {McCulloch}\ \emph {et~al.}(2013)\citenamefont
  {McCulloch}, \citenamefont {Sheludko}, \citenamefont {Junker},\ and\
  \citenamefont {Scholten}}]{ref:coldatoms2}%
  \BibitemOpen
  \bibfield  {author} {\bibinfo {author} {\bibfnamefont {A.~J.}\ \bibnamefont
  {McCulloch}}, \bibinfo {author} {\bibfnamefont {D.~V.}\ \bibnamefont
  {Sheludko}}, \bibinfo {author} {\bibfnamefont {M.}~\bibnamefont {Junker}}, \
  and\ \bibinfo {author} {\bibfnamefont {R.~E.}\ \bibnamefont {Scholten}},\
  }\href {http://dx.doi.org/10.1038/ncomms2699} {\bibfield  {journal} {\bibinfo
   {journal} {Nat Commun}\ }\textbf {\bibinfo {volume} {4}},\ \bibinfo {pages}
  {1692} (\bibinfo {year} {2013})}\BibitemShut {NoStop}%
\bibitem [{\citenamefont {Maxson}\ \emph {et~al.}(2015)\citenamefont {Maxson},
  \citenamefont {Cultrera}, \citenamefont {Gulliford},\ and\ \citenamefont
  {Bazarov}}]{ref:redmte}%
  \BibitemOpen
  \bibfield  {author} {\bibinfo {author} {\bibfnamefont {J.}~\bibnamefont
  {Maxson}}, \bibinfo {author} {\bibfnamefont {L.}~\bibnamefont {Cultrera}},
  \bibinfo {author} {\bibfnamefont {C.}~\bibnamefont {Gulliford}}, \ and\
  \bibinfo {author} {\bibfnamefont {I.}~\bibnamefont {Bazarov}},\ }\href
  {\doibase http://dx.doi.org/10.1063/1.4922146} {\bibfield  {journal}
  {\bibinfo  {journal} {Applied Physics Letters}\ }\textbf {\bibinfo {volume}
  {106}},\ \bibinfo {eid} {234102} (\bibinfo {year} {2015})}\BibitemShut
  {NoStop}%
\bibitem [{\citenamefont {Cultrera}\ \emph {et~al.}(2015)\citenamefont
  {Cultrera}, \citenamefont {Karkare}, \citenamefont {Lee}, \citenamefont
  {Liu},\ and\ \citenamefont {Bazarov}}]{ref:coldcathode}%
  \BibitemOpen
  \bibfield  {author} {\bibinfo {author} {\bibfnamefont {L.}~\bibnamefont
  {Cultrera}}, \bibinfo {author} {\bibfnamefont {S.}~\bibnamefont {Karkare}},
  \bibinfo {author} {\bibfnamefont {H.}~\bibnamefont {Lee}}, \bibinfo {author}
  {\bibfnamefont {X.}~\bibnamefont {Liu}}, \ and\ \bibinfo {author}
  {\bibfnamefont {I.}~\bibnamefont {Bazarov}},\ }\href
  {http://arxiv.org/abs/1504.05920} {\enquote {\bibinfo {title} {Cold electron
  beams from cryo-cooled, alkali antimonide photocathodes},}\ }\bibinfo
  {howpublished} {\url{http://arxiv.org/abs/1504.05920}} (\bibinfo {year}
  {2015})\BibitemShut {NoStop}%
\bibitem [{\citenamefont {Musumeci}\ \emph {et~al.}(2008)\citenamefont
  {Musumeci}, \citenamefont {Moody}, \citenamefont {England}, \citenamefont
  {Rosenzweig},\ and\ \citenamefont {Tran}}]{ref:pietro0}%
  \BibitemOpen
  \bibfield  {author} {\bibinfo {author} {\bibfnamefont {P.}~\bibnamefont
  {Musumeci}}, \bibinfo {author} {\bibfnamefont {J.~T.}\ \bibnamefont {Moody}},
  \bibinfo {author} {\bibfnamefont {R.~J.}\ \bibnamefont {England}}, \bibinfo
  {author} {\bibfnamefont {J.~B.}\ \bibnamefont {Rosenzweig}}, \ and\ \bibinfo
  {author} {\bibfnamefont {T.}~\bibnamefont {Tran}},\ }\href {\doibase
  10.1103/PhysRevLett.100.244801} {\bibfield  {journal} {\bibinfo  {journal}
  {Phys. Rev. Lett.}\ }\textbf {\bibinfo {volume} {100}},\ \bibinfo {pages}
  {244801} (\bibinfo {year} {2008})}\BibitemShut {NoStop}%
\bibitem [{\citenamefont {Gulliford}\ \emph {et~al.}(2013)\citenamefont
  {Gulliford}, \citenamefont {Bartnik}, \citenamefont {Bazarov}, \citenamefont
  {Cultrera}, \citenamefont {Dobbins}, \citenamefont {Dunham}, \citenamefont
  {Gonzalez}, \citenamefont {Karkare}, \citenamefont {Lee}, \citenamefont {Li},
  \citenamefont {Li}, \citenamefont {Liu}, \citenamefont {Maxson},
  \citenamefont {Nguyen}, \citenamefont {Smolenski},\ and\ \citenamefont
  {Zhao}}]{ref:lowemitter}%
  \BibitemOpen
  \bibfield  {author} {\bibinfo {author} {\bibfnamefont {C.}~\bibnamefont
  {Gulliford}}, \bibinfo {author} {\bibfnamefont {A.}~\bibnamefont {Bartnik}},
  \bibinfo {author} {\bibfnamefont {I.}~\bibnamefont {Bazarov}}, \bibinfo
  {author} {\bibfnamefont {L.}~\bibnamefont {Cultrera}}, \bibinfo {author}
  {\bibfnamefont {J.}~\bibnamefont {Dobbins}}, \bibinfo {author} {\bibfnamefont
  {B.}~\bibnamefont {Dunham}}, \bibinfo {author} {\bibfnamefont
  {F.}~\bibnamefont {Gonzalez}}, \bibinfo {author} {\bibfnamefont
  {S.}~\bibnamefont {Karkare}}, \bibinfo {author} {\bibfnamefont
  {H.}~\bibnamefont {Lee}}, \bibinfo {author} {\bibfnamefont {H.}~\bibnamefont
  {Li}}, \bibinfo {author} {\bibfnamefont {Y.}~\bibnamefont {Li}}, \bibinfo
  {author} {\bibfnamefont {X.}~\bibnamefont {Liu}}, \bibinfo {author}
  {\bibfnamefont {J.}~\bibnamefont {Maxson}}, \bibinfo {author} {\bibfnamefont
  {C.}~\bibnamefont {Nguyen}}, \bibinfo {author} {\bibfnamefont
  {K.}~\bibnamefont {Smolenski}}, \ and\ \bibinfo {author} {\bibfnamefont
  {Z.}~\bibnamefont {Zhao}},\ }\href {\doibase 10.1103/PhysRevSTAB.16.073401}
  {\bibfield  {journal} {\bibinfo  {journal} {Phys. Rev. ST Accel. Beams}\
  }\textbf {\bibinfo {volume} {16}},\ \bibinfo {pages} {073401} (\bibinfo
  {year} {2013})}\BibitemShut {NoStop}%
\bibitem [{\citenamefont {Gulliford}\ \emph
  {et~al.}(2015{\natexlab{a}})\citenamefont {Gulliford}, \citenamefont
  {Bartnik}, \citenamefont {Bazarov}, \citenamefont {Dunham},\ and\
  \citenamefont {Cultrera}}]{ref:lowemitter2}%
  \BibitemOpen
  \bibfield  {author} {\bibinfo {author} {\bibfnamefont {C.}~\bibnamefont
  {Gulliford}}, \bibinfo {author} {\bibfnamefont {A.}~\bibnamefont {Bartnik}},
  \bibinfo {author} {\bibfnamefont {I.}~\bibnamefont {Bazarov}}, \bibinfo
  {author} {\bibfnamefont {B.}~\bibnamefont {Dunham}}, \ and\ \bibinfo {author}
  {\bibfnamefont {L.}~\bibnamefont {Cultrera}},\ }\href {\doibase
  http://dx.doi.org/10.1063/1.4913678} {\bibfield  {journal} {\bibinfo
  {journal} {Applied Physics Letters}\ }\textbf {\bibinfo {volume} {106}},\
  \bibinfo {eid} {094101} (\bibinfo {year} {2015}{\natexlab{a}})}\BibitemShut
  {NoStop}%
\bibitem [{\citenamefont {Bartnik}\ \emph {et~al.}(2015)\citenamefont
  {Bartnik}, \citenamefont {Gulliford}, \citenamefont {Bazarov}, \citenamefont
  {Cultera},\ and\ \citenamefont {Dunham}}]{ref:lowemitter3}%
  \BibitemOpen
  \bibfield  {author} {\bibinfo {author} {\bibfnamefont {A.}~\bibnamefont
  {Bartnik}}, \bibinfo {author} {\bibfnamefont {C.}~\bibnamefont {Gulliford}},
  \bibinfo {author} {\bibfnamefont {I.}~\bibnamefont {Bazarov}}, \bibinfo
  {author} {\bibfnamefont {L.}~\bibnamefont {Cultera}}, \ and\ \bibinfo
  {author} {\bibfnamefont {B.}~\bibnamefont {Dunham}},\ }\href {\doibase
  10.1103/PhysRevSTAB.18.083401} {\bibfield  {journal} {\bibinfo  {journal}
  {Phys. Rev. ST Accel. Beams}\ }\textbf {\bibinfo {volume} {18}},\ \bibinfo
  {pages} {083401} (\bibinfo {year} {2015})}\BibitemShut {NoStop}%
\bibitem [{\citenamefont {Gulliford}\ \emph
  {et~al.}(2015{\natexlab{b}})\citenamefont {Gulliford}, \citenamefont
  {Bartnik},\ and\ \citenamefont {Bazarov}}]{ref:coldgun}%
  \BibitemOpen
  \bibfield  {author} {\bibinfo {author} {\bibfnamefont {C.}~\bibnamefont
  {Gulliford}}, \bibinfo {author} {\bibfnamefont {A.}~\bibnamefont {Bartnik}},
  \ and\ \bibinfo {author} {\bibfnamefont {I.}~\bibnamefont {Bazarov}},\ }\href
  {http://arxiv.org/abs/1510.07738} {\enquote {\bibinfo {title} {Cold electron
  beams from cryo-cooled, alkali antimonide photocathodes},}\ }\bibinfo
  {howpublished} {\url{http://arxiv.org/abs/1510.07738}} (\bibinfo {year}
  {2015}{\natexlab{b}})\BibitemShut {NoStop}%
\bibitem [{\citenamefont {Bazarov}\ \emph {et~al.}(2009)\citenamefont
  {Bazarov}, \citenamefont {Dunham},\ and\ \citenamefont
  {Sinclair}}]{ref:maxbb}%
  \BibitemOpen
  \bibfield  {author} {\bibinfo {author} {\bibfnamefont {I.~V.}\ \bibnamefont
  {Bazarov}}, \bibinfo {author} {\bibfnamefont {B.~M.}\ \bibnamefont {Dunham}},
  \ and\ \bibinfo {author} {\bibfnamefont {C.~K.}\ \bibnamefont {Sinclair}},\
  }\href {\doibase 10.1103/PhysRevLett.102.104801} {\bibfield  {journal}
  {\bibinfo  {journal} {Phys. Rev. Lett.}\ }\textbf {\bibinfo {volume} {102}},\
  \bibinfo {pages} {104801} (\bibinfo {year} {2009})}\BibitemShut {NoStop}%
\bibitem [{\citenamefont {Filippetto}\ \emph {et~al.}(2014)\citenamefont
  {Filippetto}, \citenamefont {Musumeci}, \citenamefont {Zolotorev},\ and\
  \citenamefont {Stupakov}}]{ref:pietro2}%
  \BibitemOpen
  \bibfield  {author} {\bibinfo {author} {\bibfnamefont {D.}~\bibnamefont
  {Filippetto}}, \bibinfo {author} {\bibfnamefont {P.}~\bibnamefont
  {Musumeci}}, \bibinfo {author} {\bibfnamefont {M.}~\bibnamefont {Zolotorev}},
  \ and\ \bibinfo {author} {\bibfnamefont {G.}~\bibnamefont {Stupakov}},\
  }\href {\doibase 10.1103/PhysRevSTAB.17.024201} {\bibfield  {journal}
  {\bibinfo  {journal} {Phys. Rev. ST Accel. Beams}\ }\textbf {\bibinfo
  {volume} {17}},\ \bibinfo {pages} {024201} (\bibinfo {year}
  {2014})}\BibitemShut {NoStop}%
\bibitem [{\citenamefont {Li}\ \emph {et~al.}(2012)\citenamefont {Li},
  \citenamefont {Roberts}, \citenamefont {Scoby}, \citenamefont {To},\ and\
  \citenamefont {Musumeci}}]{ref:pietro1}%
  \BibitemOpen
  \bibfield  {author} {\bibinfo {author} {\bibfnamefont {R.~K.}\ \bibnamefont
  {Li}}, \bibinfo {author} {\bibfnamefont {K.~G.}\ \bibnamefont {Roberts}},
  \bibinfo {author} {\bibfnamefont {C.~M.}\ \bibnamefont {Scoby}}, \bibinfo
  {author} {\bibfnamefont {H.}~\bibnamefont {To}}, \ and\ \bibinfo {author}
  {\bibfnamefont {P.}~\bibnamefont {Musumeci}},\ }\href {\doibase
  10.1103/PhysRevSTAB.15.090702} {\bibfield  {journal} {\bibinfo  {journal}
  {Phys. Rev. ST Accel. Beams}\ }\textbf {\bibinfo {volume} {15}},\ \bibinfo
  {pages} {090702} (\bibinfo {year} {2012})}\BibitemShut {NoStop}%
\bibitem [{\citenamefont {Han}(2011)}]{ref:rfsimpap}%
  \BibitemOpen
  \bibfield  {author} {\bibinfo {author} {\bibfnamefont {J.-H.}\ \bibnamefont
  {Han}},\ }\href {\doibase 10.1103/PhysRevSTAB.14.050101} {\bibfield
  {journal} {\bibinfo  {journal} {Phys. Rev. ST Accel. Beams}\ }\textbf
  {\bibinfo {volume} {14}},\ \bibinfo {pages} {050101} (\bibinfo {year}
  {2011})}\BibitemShut {NoStop}%
\bibitem [{\citenamefont {Neal}(1968)}]{ref:slaccav}%
  \BibitemOpen
  \bibfield  {author} {\bibinfo {author} {\bibfnamefont {R.}~\bibnamefont
  {Neal}},\ }\href {https://books.google.com/books?id=ip8VAQAAIAAJ} {\emph
  {\bibinfo {title} {The Stanford two-mile accelerator}}},\ \bibinfo {number}
  {v. 1}\ (\bibinfo  {publisher} {W. A. Be},\ \bibinfo {year}
  {1968})\BibitemShut {NoStop}%
\bibitem [{\citenamefont {Bazarov}\ \emph {et~al.}(2011)\citenamefont
  {Bazarov}, \citenamefont {Kim}, \citenamefont {Lakshmanan},\ and\
  \citenamefont {Maxson}}]{ref:dcrfcomp}%
  \BibitemOpen
  \bibfield  {author} {\bibinfo {author} {\bibfnamefont {I.~V.}\ \bibnamefont
  {Bazarov}}, \bibinfo {author} {\bibfnamefont {A.}~\bibnamefont {Kim}},
  \bibinfo {author} {\bibfnamefont {M.~N.}\ \bibnamefont {Lakshmanan}}, \ and\
  \bibinfo {author} {\bibfnamefont {J.~M.}\ \bibnamefont {Maxson}},\ }\href
  {\doibase 10.1103/PhysRevSTAB.14.072001} {\bibfield  {journal} {\bibinfo
  {journal} {Phys. Rev. ST Accel. Beams}\ }\textbf {\bibinfo {volume} {14}},\
  \bibinfo {pages} {072001} (\bibinfo {year} {2011})}\BibitemShut {NoStop}%
\bibitem [{\citenamefont {Maxson}\ \emph {et~al.}(2013)\citenamefont {Maxson},
  \citenamefont {Bazarov}, \citenamefont {Wan}, \citenamefont {Padmore},\ and\
  \citenamefont {Coleman-Smith}}]{ref:DIH}%
  \BibitemOpen
  \bibfield  {author} {\bibinfo {author} {\bibfnamefont {J.~M.}\ \bibnamefont
  {Maxson}}, \bibinfo {author} {\bibfnamefont {I.~V.}\ \bibnamefont {Bazarov}},
  \bibinfo {author} {\bibfnamefont {W.}~\bibnamefont {Wan}}, \bibinfo {author}
  {\bibfnamefont {H.~A.}\ \bibnamefont {Padmore}}, \ and\ \bibinfo {author}
  {\bibfnamefont {C.~E.}\ \bibnamefont {Coleman-Smith}},\ }\href
  {http://stacks.iop.org/1367-2630/15/i=10/a=103024} {\bibfield  {journal}
  {\bibinfo  {journal} {New Journal of Physics}\ }\textbf {\bibinfo {volume}
  {15}},\ \bibinfo {pages} {103024} (\bibinfo {year} {2013})}\BibitemShut
  {NoStop}%
\end{thebibliography}
\end{document}